\documentclass[twocolumn]{aastex701} 

\usepackage{multirow}

\usepackage{graphics,epsf}
\usepackage[utf8]{inputenc}
\usepackage{amsmath}                % American Mathematical Society package
\usepackage{amsfonts}               % American Mathematical Society fonts
\usepackage{amssymb}                % American Mathematical Society symbol
\usepackage{epsfig}                 % EPS figures
\usepackage{graphicx}               % Required for inserting images
\usepackage{float}
\usepackage{color}
\usepackage{multirow}               % double row table entries

\hypersetup{
    colorlinks=true,
    linkcolor=red,   
    urlcolor=cyan}

\usepackage[colorinlistoftodos]{todonotes}

\newcommand{\km}{{~\rm km}}
\newcommand{\s}{{~\rm s}}

\newcommand{\g}{{~\rm g}}
\newcommand{\G}{{~\rm G}}
\newcommand{\K}{{~\rm K}}

\begin{document}
\title{Supernova remnant 0509-67.5 is consistent with an explosion inside an old planetary nebula (SNIP)}
%\date{September 2024}

\author[0000-0003-0375-8987]{Noam Soker}
\affiliation{Department of Physics, Technion - Israel Institute of Technology, Haifa, 3200003, Israel; soker@technion.ac.il}
\email{soker@technion.ac.il}

\begin{abstract}
I critically examine claims in the paper \href{https://arxiv.org/abs/2608.11978}{arXiv:2608.11978}  that the double-detonation (DDet) scenario explains the type Ia supernova (SN Ia) remnant (SNR Ia) SNR~0509-67.5, and find the arguments supporting the DDet scenario weak; hence, I reiterate my claim that the core-degenerate (CD) scenario, where a lonely white dwarf (WD), which is the merger product of a lower-mass WD and the core of an asymptotic giant branch star,  exploded inside an old planetary nebula, i.e., an SNIP, best explains  SNR~0509-67.5. I find that the flat edge of the SNR in the north-northeast, which in the DDet scenario is due to a shadow by the companion to the WD that exploded, is not unique in this SNR, and that the circumstellar matter, i.e., an old planetary nebula, shaped this edge, as well as other structures on the edge of this SNR. I emphasize that analyses of SNRs Ia and SNe Ia at late stages must consider the claim that most normal SNe Ia are SNIPs, implying that old planetary nebulae can heavily shape their morphologies. The bulk velocity inferred from an iron emission line in SNR~0509-67.5, which the DDet explosion attributes to the orbital motion of the exploding WD around its companion, can be an outcome of the explosion of a near-Chandrasekhar lonely WD; an off-center delayed-detonation transition explosion can form asymmetrical nucleosynthesis, while leading to a bulk motion of nickel that decays to iron. I repeat my claim that the CD scenario of a SNIP is the most likely scenario for SNR~0509-67.5. 
\end{abstract}


% ==============================================
\section{INTRODUCTION}
\label{sec:intro}
% ==============================================

The large number of reviews of normal and peculiar type Ia supernovae (SNe Ia) in the last decade reflects the disagreement on the dominant scenarios of normal and peculiar SNe Ia, the way to classify them, and the interpretation of some observations (\citealt{MaedaTerada2016, Hoeflich2017, LivioMazzali2018, Soker2018Rev, Soker2019Rev, Soker2024Rev, Wang2018,  Jhaetal2019NatAs, RuizLapuente2019, Ruiter2020, Aleoetal2023, Liuetal2023Rev, Vinkoetal2023, RuiterSeitenzahl2025}). There are advantages to all SN Ia scenarios, but, on the other hand, all SN Ia scenarios encounter challenges in explaining some observations, overcoming theoretical difficulties, or both (e.g., \citealt{Pearsonetal2024, Sharonetal2024, Sharonetal2025a, SchinasiLembergKushnir2025, Sharonetal2025b, Wangetal2024, Glanzetal2025, MengPodsiadlowski2026, Glanzetal2027}; see \citealt{Soker2026SNIPs} for a list of the scenarios).  
      
In addition to the general dispute over the dominant SN Ia scenarios, there are some disputes on particular SNe Ia or SN Ia remnants (SNRs Ia).  
Such is SNR~0509-67.5 (J0509-6731), an SNR Ia in the Large Magellanic Cloud (LMC): some claim the double-detonation (DDet) scenario \citep{Dasetal2025, Dasetal2026, Mandaletal2026}, while I claim the core-degenerate (CD) scenario \citep{Soker2025SNR0509}. 
\cite{Dasetal2025} explained the double-shell segments of SNR~0509-67.5 with the DDet scenario. 
In \cite{Soker2025SNR0509}, I argued that the similar iron and hydrogen elliptical morphologies of SNR 0509-67.5 hint that the ambient gas shaped the ejecta; this ambient gas is a remnant of
an old planetary nebula. Namely, SNR 0509-67.5 is a SN Ia inside a planetary nebula, a SNIP. Given that no stellar survivor has been observed (e.g., \citealt{SchaeferPagnotta2012, Panetal2014, Notaetal2016, Shieldsetal2023}), I argued there that the most likely scenario is the CD scenario. 

In a paper posted recently on the arXiv (\href{https://arxiv.org/abs/2608.11978}{arXiv:2608.11978}), \cite{Dasetal2026} further argue for the DDet scenario, and not a SNIP. I critically examine their claims and reiterate my claim that SNR 0509-67.5 is a SNIP.\footnote{The paper was posted after publication in MNRAS. I think it is time journal editors forced authors to post their papers on arXiv for at least a week before final acceptance. This would allow the community to comment. The only publication immune to corrections is the Bible, and it should stay like this.} 

% ===========================================
\section{The flat edge}
\label{sec:FlatEdge}
% ===========================================

\begin{figure*}[t]
%\begin{center}
\includegraphics[trim=0.0cm 22.9cm 25.0cm 0.0cm,scale=1.00]{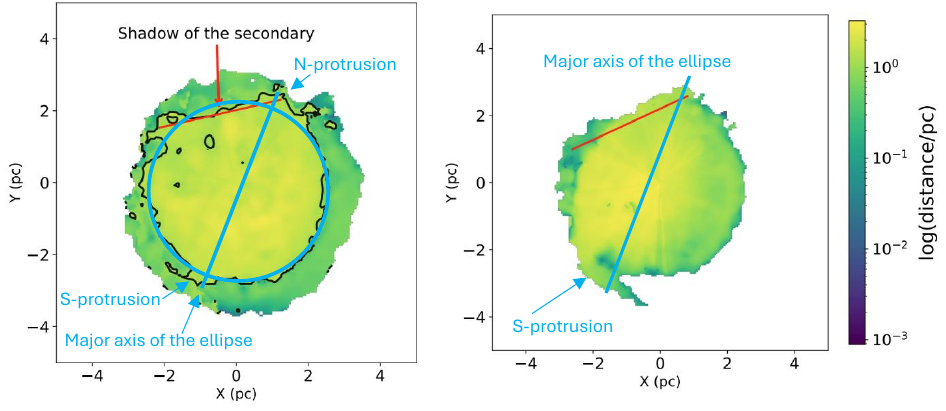} %[trim=left lower right upper]
\caption{A figure adapted from \cite{Dasetal2026}, with my marks in pale blue. The major axis of the ellipse is the one \cite{Soker2025SNR0509} defined, panel (d) in Figure \ref{Fig:Fe} here. 
\cite{Dasetal2026} marked the flat edge with a red line and argued that it is the shadow of the WD companion to the WD that exploded. I attribute the two opposite protrusions to shaping by the planetary nebula into which SNR 0509-67.5 exploded. 
Left: The highly red-shifted ejecta.  Right: The blue-shifted ejecta. Note that the two red lines are $12^\circ$ apart. 
}
\label{Fig:FlatEdge}
%\end{center}
\end{figure*}

\begin{figure*}[t]
\begin{center}
\includegraphics[trim=0.0cm 10.5cm 0.0cm 0.0cm,scale=0.81]{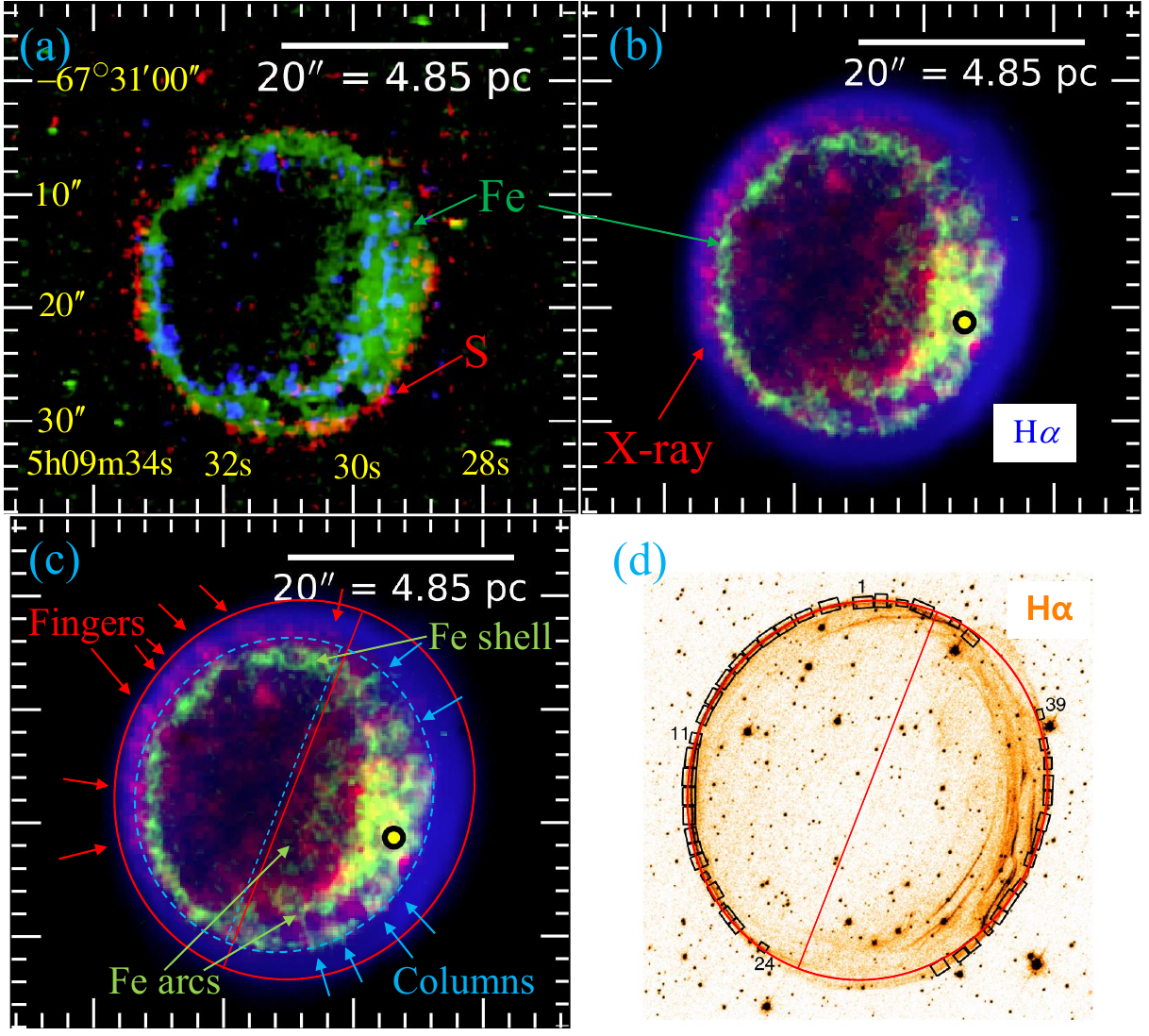} %[trim=left lower right upper]
\caption{ A figures from \cite{Soker2025SNR0509} of images of SNR~0509-67.5. Panels a-c are composite images adapted from \cite{Seitenzahletal2019}. (a) Red: [S \textsc{xii}] $7613.1 \text{\AA}$; green: [Fe \textsc{xiv}] $5303 \text{\AA}$; blue: [Fe \textsc{ix}] $8236.8 \text{\AA}$. (b+c) Red: an X-ray image from Chandra ACIS (for an X-ray image, see also, e.g., \citealt{WarrenHughes2004}); green: [Fe \textsc{xiv}] VLT-MUSE); blue: H$\alpha$ (VLT-MUSE). (Yellow dot: region of spectra extraction.) In panel c, I added relevant marks. The red ellipse is as in panel d, with its long axis (the straight red line) 1.075 times its short axis. The pale-blue dashed ellipse is 0.83 times that of the red ellipse and has the same orientation. The pale-blue dashed ellipse is displaced to mark the edge of the [Fe \textsc{xiv}] region. \cite{Soker2025SNR0509} attributed the X-ray fingers and Fe columns to Rayleigh-Taylor instabilities. (d) An H$\alpha$ image (ACS-HST) adapted from \cite{Hoveyetal2015}. \cite{Soker2025SNR0509} added the red ellipse through the rectangles on the exterior H$\alpha$ rim that \cite{Hoveyetal2015} drew.  } 
\label{Fig:Fe}
\end{center}
\end{figure*}

\cite{Dasetal2026} presented the red- and blue-shifted ejecta according to the  [Fe \textsc{xiv}]  $\lambda5303$ emission line. I present two panels from their figure in Figure \ref{Fig:FlatEdge}. They argue that the flat edge, which they marked with straight red lines on the two panels (they are at $12^\circ$ to each other), is due to a shadow by the companion to the exploding white dwarf (WD).  I drew a circle in the left panel and a straight line through the center of the circle, which is as the orientation of the major axis of the ellipse through the bright H$\alpha$ emission, as I defined in \cite{Soker2025SNR0509}, and show it here in Figure \ref{Fig:Fe}.  I note the two opposite protrusions, which I attribute to shaping by two opposite structural features in the planetary nebula into which SNR 0509-67.5 exploded (see \citealt{TsebrenkoSoker2013} for simulations of SNIPs that obtain such `ears'). 
% FFFFFFFFFFFFFFFFFFFFFFFFFFFFFFFFFFFFFFFFFF
% FFFFFFFFFFFFFFFFFFFFFFFFFFFFFFFFFFFFFFFFFF
% FFFFFFFFFFFFFFFFFFFFFFFFFFFFFFFFFFFFFFFFFF
% FFFFFFFFFFFFFFFFFFFFFFFFFFFFFFFFFFFFFFFFFF

\begin{figure}[t]
\begin{center}
\includegraphics[trim=0.0cm 20.cm 0.0cm 0.0cm,scale=0.81]{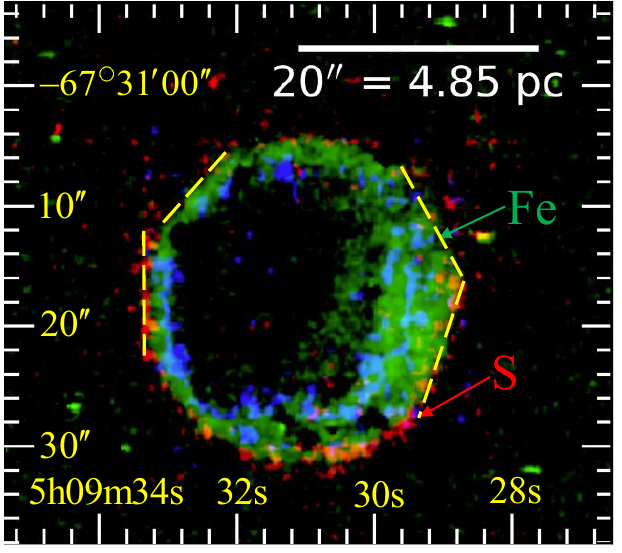} %[trim=left lower right upper]
\caption{Like panel (a) of Figure \ref{Fig:Fe}, but with dashed-orange lines, I mark four possible flat edges according to the S or Fe structure. The flat edge that \cite{Dasetal2026} argued for is not unique.  } 
\label{Fig:MoreEdges}
\end{center}
\end{figure}

\begin{figure}[t]
\begin{center}
\includegraphics[trim=0.0cm 14.8cm 0.0cm 0.0cm,scale=0.53]{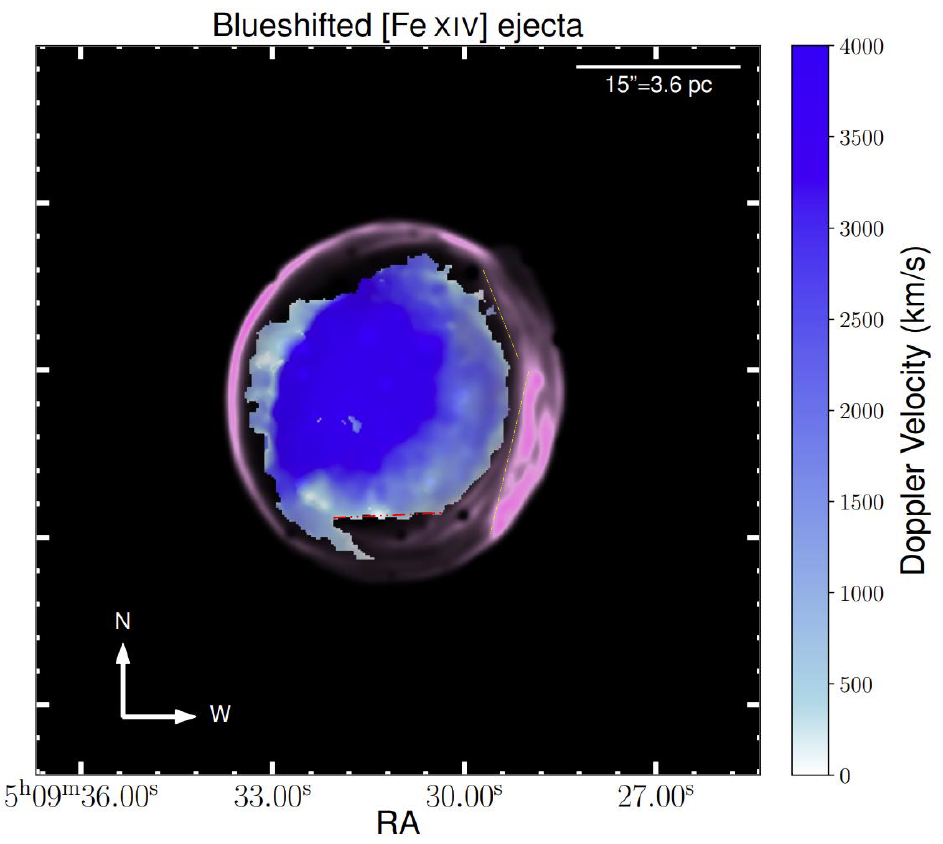} %[trim=left lower right upper]
\caption{An image adapted from \cite{Dasetal2026} of the blue-shifted component of the Doppler map of SNR~0509-67.5 in the [Fe \textsc{xiv}] emission line. The H$\alpha$ emission is overlaid in magenta. I added three flat structures: two (in dotted-dotted-dashed yellow line) of the hydrogen-rich ambient gas, which I consider to be an old planetary nebula, and one of the ejecta (red line). }
\label{Fig:DopplerEdges}
\end{center}
\end{figure}

I dispute the claim by \cite{Dasetal2026} that the flat edge is due to a shadow cast by a companion for the following reasons. 
\begin{enumerate}
\item Although it is seen in the blue- and red-shifted images that \cite{Dasetal2026} presented (Figure \ref{Fig:FlatEdge} here), it is not seen in other images, like those in Figure \ref{Fig:Fe}.   
\item The side of the flat edge, north-northeast, has the most prominent structure of fingers (panel c of Figure \ref{Fig:Fe}). I attributed \citep{Soker2025SNR0509} the fingers to Rayleigh-Taylor instabilities resulting from the ejecta interaction with the CSM (the old planetary nebula, as SNR 0509 is a SNIP; \citealt{TsebrenkoSoker2013} obtain such fingers in their SNIP simulations). I suggest that the flat edge results from the interaction of the ejecta with the CSM.    
\item In the DDet scenario, there is no structural feature on the opposite side of the flat edge caused by the shadow of the companion. However, SNR~0509-67.5 possesses a reach and prominent iron arcs in the south-southwest, on the opposite side of the flat edge (panel c of Figure \ref{Fig:Fe}). This might suggest some opposite structural features in the CSM that caused the two opposite structures, although they are different.  
\item The S and Fe structures exhibit more than one flat edge, as evident from panel (a) of Figure \ref{Fig:Fe}. I drew the same figure but with the marks of the four flat edges (dashed orange lines) in Figure \ref{Fig:MoreEdges}. The flat edge that \cite{Dasetal2026} argued for is not unique. In Figure \ref{Fig:DopplerEdges}, I present another image from \cite{Dasetal2026}. This image includes H$\alpha$ emission map, on which I identify two flat inner edges, as I marked with dotted-dotted-dashed yellow lines. With a dotted-dotted-dashed red line, I marked a southern flat edge in the ejecta.  
\end{enumerate}
% FFFFFFFFFFFFFFFFFFFFFFFFFFFFFFFFFFFFFFFFFF
% FFFFFFFFFFFFFFFFFFFFFFFFFFFFFFFFFFFFFFFFFF
% FFFFFFFFFFFFFFFFFFFFFFFFFFFFFFFFFFFFFFFFFF
% FFFFFFFFFFFFFFFFFFFFFFFFFFFFFFFFFFFFFFFFFF

The discussion in this section shows that the arguments \cite{Dasetal2026} presented for the flat edge being the result of the DDet scenario are weak. I prefer the CD scenario, as SNR~0509-67.5 is probably an SNIP, where an old planetary nebula (hydrogen-rich ambient gas) shaped the flat edges.   

% ==========================================
\section{the bulk velocity}
\label{sec:bulk}

% ==========================================

\cite{Dasetal2026} measured the bulk Doppler velocity of the [Fe \textsc{xiv}]  $\lambda5303$ emission line to be $-1000 \pm 60 \km \s^{-1}$, which they interpreted as the line-of-sight component of the exploding WD orbital velocity in the framework of the DDet scenario. 

I note that off-center delayed-detonation transition (deflagration to detonation transition) in near-Chandra-mass WD explosions (as the CD scenario assumes) can lead to asymmetrical explosions with polarization and asymmetrical velocities of metals (e.g., \citealt{Hoeflichetal2023, Cikotaetal2026}), with velocity differences between opposite sides of $\approx 1000 \km \s^{-1}$ (e.g., \citealt{Shiberetal2026a, Shiberetal2026b} for recent studies). \cite{Hoeflichetal2021} convincingly demonstrate bulk non-zero velocity in off-center delayed-detonation transition explosions in their simulations and comparison to SN 2020qxp (see their figure 3). Note that the continuum polarization of normal SNe Ia is generally very low (e.g., \citealt{Yangetal2020})

I argue that the bulk velocity measured in the iron line results from an asymmetrical nucleosynthesis of $^{56}$Ni during an explosion of a near-Chandrasekhar mass WD, in the framework of the CD scenario.

% ==========================================
\section{Summary}
\label{sec:Summary}
% ==========================================

This paper is another in the dispute over the scenario of SNR~0509-67.5: whether the DDet scenario (e.g., \citealt{Dasetal2025, Mandaletal2026}) or the CD scenario (e.g., \citealt{Soker2025SNR0509}). I analyzed recent claims by \cite{Dasetal2026}
where they further argued for the DDet scenario.  
 
In Section \ref{sec:FlatEdge}, I listed four reasons why the flat edge they identified is unlikely to be a shadow by the companion to the WD that exploded. I argue that the structures on the edge of this SNR are due to an interaction with an old planetary nebula, as SNR~0509-67.5 is an SNIP \citep{Soker2025SNR0509}. 

The protrusions that I marked in Figure \ref{Fig:FlatEdge}, according to the SNIP scenario, are from the planetary nebula phase. The SN Ia ejecta interacts with the planetary nebula, and the two protrusions (also termed ears) survive, as \cite{TsebrenkoSoker2013} showed in their hydrodynamical simulations. \cite{TsebrenkoSoker2015} simulated SNR G1.9+0.3 as an SNIP, and showed that more complicated structures in the planetary nebula, such as clumps, can lead to more complicated structures in the SNR Ia after the ejecta interacts with the planetary nebula; I later strengthened the case that SNR G1.9+0.3 is an SNIP \citep{Soker2024RAA}. Such  an interaction might account for the complicated arcs on the southwest in SNR~0509-67.5

In general, studies that analyze SNRs Ia and SNe Ia at late stages should consider the claim that most ($\simeq 70-90\%$) normal SNe Ia are SNIPs (e.g., \citealt{Soker2026SNIPs}); I take the recent finding of an AGB progenitor of an SN Ia-CSM by  \cite{Szalaietal2026} to strengthen this claim. Namely, many features and properties of normal SNe Ia and SNRs Ia can result from the interaction of the ejecta with an old planetary nebula. 

\cite{Dasetal2026} measured a bulk velocity of $-1000 \pm 60 \km \s^{-1}$ in the [Fe \textsc{xiv}] $\lambda5303$ emission line, and attributed it to the orbital velocity of the exploding WD just before the explosion. In Section \ref{sec:bulk}, I argue that this bulk velocity can result from an off-center delayed-detonation transition explosion. 

Overall, I find the arguments of \cite{Dasetal2026} for the DDet scenario of SNR~0509-67.5 to be weak. I prefer the CD scenario of an SNIP, as I argued for in \cite{Soker2025SNR0509}.

% ===============================
% \section*{Acknowledgments}
% ===============================

%%%%%%%%%%%%%%%%%%%%%%%%%%%

%%%%%%%%%%%%%%%%%%%%%%%%%%%

% %%%%%%%%%%%%  References %%%%%%%%%%%%%%%%%%%%%

%\newpage 

\end{document}